\documentclass[twocolumn,amsmath,amssymb,amssymb, showpacs, aps, pra]{revtex4}
\usepackage{graphicx}
\usepackage{dcolumn}
\usepackage{bm}
\usepackage{braket}
\usepackage{mathrsfs}
\begin{document}
\title{Analytical solution to position dependent mass Schr$\ddot{\text{o}}$dinger equation}
\author{Pankaj K. Jha$^{1,*}$, Hichem Eleuch$^{1,2}$, Yuri V. Rostovtsev$^{1,3}$}
\affiliation{$^{1}$Institute for Quantum Science and Engineering and Department of Physics and Astronomy, Texas A\&M University, College Station, Texas 77843, USA\\
$^{2}$Physics and Astronomy Department, College of Science, King Saud University, Riyadh 11451, Saudi Arabia\\
$^{3}$Department of Physics, University of North Texas, Denton, Texas 76203, USA}
\begin{abstract}
\noindent Using a recently developed technique to solve Schr$\ddot{\text{o}}$dinger equation for constant mass, we studied the regime in which mass varies with position i.e position dependent mass Schr$\ddot{\text{o}}$dinger equation(PDMSE). We obtained an analytical solution for the PDMSE and applied our approach to study a position dependent mass $m(x)$ particle scattered by a potential $\mathcal{V}(x)$. We also studied the structural analogy between PDMSE and two-level atomic system interacting with a classical field.
\end{abstract}
\maketitle
Schr$\ddot{o}$dinger equation with position dependent mass is one of the areas of research which has gained great attention in the past decades. Position-dependent mass Schr$\ddot{o}$dinger equation(PDMSE) has been applied to several physical systems. For example PDMSE is applied in electronic properties of semiconductors \cite{I1}, quantum dots and quantum wells~\cite{I2,I3}, semiconductors hetero-structures~\cite{I4}, supper-lattice band structures~\cite{I5}, He-Clusters~\cite{I6} quantum liquids~\cite{I7}, the dependence of energy gap on magnetic field in semiconductor nano-scale rings~\cite{I8}, the solid state problem with Dirac equation~\cite{I9} etc.
One of the motivation behind investigating these system with position dependent mass is, how do these mass variation effects the dynamics of the quantum-mechanical system. In such systems the energy is described by a Hamiltonian which contains the kinetic energy and the potential energy opeators $\mathscr{H}=\mathscr{T}+\mathscr{V}$, where special care is taken for the kinetic term.

O. von Roos~\cite{I10} was the first to suggest the following generalized form of the kinetic energy operator for position-dependent mass model
\begin{equation}\label{50}
\mathscr{T}=\frac{1}{4}(m^{\eta}\mbox{\bf{p}}m^{\epsilon}\mbox{\bf{p}}m^{\rho}+m^{\rho}\mbox{\bf{p}}m^{\epsilon}\mbox{\bf{p}}m^{\eta}),
\end{equation}
where $m=m(\bf{r})$ is the position-dependent mass. The constants $\eta,\epsilon$ and $\rho $, which are also knows as the von Roos ambiguity parameters can be assumed to be arbitrary  but they obey the constraint equation $\eta +\epsilon +\rho = -1$. Before von Roos several forms of operator $\mathscr{T}$ has been used to solve this problem~\cite{I11,I12,I13,I14}. Different approaches have been used to find analytical solution to PDMSE like point-canonical transformation~\cite{S1}, Green's function~\cite{S2}, Heun equation~\cite{S3,S4}, Group-Theoretical method~\cite{S5}, potential algebra~\cite{S6}, Lie-algebraic~\cite{S7} and supersymmetric~\cite{S8} approach etc.

In this Brief Report, we obtained an analytical solution for the position dependent mass Schr$\ddot{\text{o}}$dinger equation with general mass variation $m(x)$ by transforming the PDMSE to Riccati equation. Analytical solution for Schr$\ddot{\text{o}}$dinger equation with constant mass, beyond adiabatic approximation, has been recently investigated extensively~\cite{I15} where we applied our approach to 1-D \cite{I15} and 3-D \cite{I15b} scattering problem and showed that our method gives better accuracy than the well-known JWKB method. Here in this article we have extended that approach to variable mass regime and obtained very accurate results for PDMSE. The main result of this paper is the analytical solution given by Eq.(\ref{518}). To illustrate how well our method works we considered a position dependent mass $m(x)$ particle scattered by potential $V(x)$ and obtained the wave function for 1D case (which could easily be extended to 3D). Plots of the numerical simulation and the analytical solution are nearly identical [see figs 1,2,3]. We also briefly discuss structural analogy between PDMSE and two-level atom(TLA) driven by resonant classical field thus bridging between quantum optics and condensed matter Physics.

\section{Model: Analytical Solution}
\noindent Let  us consider the symmetric ordering form of the kinetic energy operator as given by Eq.(\ref{50}) i.e $(\eta =0, \epsilon=-1,\rho = 0)$.
The Hamiltonian of the system is written as
\begin{equation} \label{51}
\mathscr{H}=\frac{1}{2}\left[\mbox{\bf{p}}\frac{1}{m}\mbox{\bf{p}}\right] + \mathscr{V}(\bf{r}).
\end{equation}
In 1-D, this Hamiltonian gives the following form of the Schr$\ddot{o}$dinger equation with position dependent mass,
\begin{equation} \label{52}
-\frac{\hbar^{2}}{2}\frac{d}{dx}\left[\frac{1}{m(x)}\frac{d\Psi(x)}{dx}\right]+\mathcal{V}(x)\Psi(x)=\mathcal{E}\Psi(x).
\end{equation}
Here the mass function $m(x)=m_{0}\varrho(x)$. Let us non dimensionlize Eq.(\ref{52}) using the following scaling parameters
\begin{equation} \label{53}
E=\mathcal{E/E}_{0}, \quad V(x)=\mathcal{V}(x)/\mathcal{E}_{0}, \quad z=\left(\sqrt{2m_{0}\mathcal{E}_{0}}/\hbar\right)x.
\end{equation}
we get,
\begin{equation} \label{54}
-\frac{d}{dz}\left[\frac{1}{\varrho(z)}\frac{d\Psi(z)}{dz}\right]+V(x)\Psi(z)=E\Psi(z).
\end{equation}
Let us recast Eq.(\ref{54}) in a more desirable form we need,
\begin{equation} \label{55}
\left\{\frac{d^{2}}{dz^{2}}-\frac{\varrho'(z)}{\varrho(z)}\frac{d}{dz}+\varrho(z)[E-V(z)]\right\}\Psi(z)=0.
\end{equation}
Let us define two functions $\zeta(z)$ and $\xi(z)$ as,
\begin{subequations} \label{56}
\begin{align}
\zeta(z)&= \frac{\varrho'(z)}{2\varrho(z)},\label{second}\\
\xi(z)&= \varrho(z)[E-V(z)].
\end{align}
\end{subequations}
Substituting Eq.(\ref{56}) in Eq.(\ref{55}) we get,
\begin{equation} \label{57}
\left\{\frac{d^{2}}{dz^{2}}-2\zeta(z)\frac{d}{dz}+\xi(z)\right\}\Psi(z)=0.
\end{equation}
To find an analytical solution for the Eq.(\ref{57}) we make a formal substitution of
\begin{equation} \label{58}
\Psi(z)=e^{\int {f(\tilde{z})d\tilde{z}}},
\end{equation}
which reduces the Schr$\ddot{o}$dinger equation to non-linear Riccati equation of the form
\begin{equation} \label{59}
f'(z)+f(z)^{2}-2\zeta(z)f(z)+\xi(z)=0.
\end{equation}
Let us work in the adiabatic regime of the problem i.e to the zeroth order and neglect the contribution of $f'(z)$ in Eq.(\ref{59}).
Solving Eq.(\ref{59}) for $f'(z)=0$ we get,
\begin{equation} \label{510}
f_{0}(z)=\zeta(z)\pm i\sqrt{\xi(z)-\zeta(z)^{2}}.
\end{equation}
The zeroth order general solution of the Schr$\ddot{o}$dinger equation with position dependent mass Eq.(\ref{57}) is
\begin{equation} \label{511}
\Psi(z)=A_{1}e^{\varphi(z)+i\theta(z)}+A_{2}e^{\varphi(z)-i\theta(z)},
\end{equation}
where,
\begin{subequations}\label{512}
\begin{align}
\varphi(z)&= \int^{z}_{z_{0}}{\zeta(\tilde{z})}d\tilde{z}\label{second},\\
\theta(z)&=  \int^{z}_{z_{0}}{\sqrt{\xi(\tilde{z})-\zeta(\tilde{z})^{2}}} d\tilde{z}.
\end{align}
\end{subequations}
To go beyond the zeroth order approximation assume,
\begin{equation} \label{513}
f_{1}(z)=f_{0}(z)+\epsilon_{1}(z).
\end{equation}
Substituting Eq.(\ref{513}) in Eq.(\ref{59}) and neglecting the term $\propto$ $\epsilon^{2}(z) $, Eq.(\ref{59}) gives
\begin{equation} \label{514}
f'_{0}(z)+\epsilon_{1}'(z)+2f_{0}(z)\epsilon_{1}(z)-2\zeta(z)\epsilon_{1}(z)=0.
\end{equation}
The general solution to Eq(\ref{514}) is given as
\begin{equation} \label{515}
\epsilon_{1}(z)=\left\{\int^{z}_{z_{0}}\left[ -f'_{0}(\tilde{z})e^{-u(\tilde{z})} \right]d\tilde{z}+{\cal B}\right\}e^{u(z)},
\end{equation}
where,
\begin{equation} \label{516}
u(\tilde{z})=2\int^{\tilde{z}}_{z_{0}}[-f_{0}(z'')+\zeta(z'')]dz''.
\end{equation}
Let us consider $\epsilon_{1}(z_{0})\approx 0$ which gives ${\cal B}=0$ hence
 \begin{equation} \label{517}
\epsilon_{1}(z)=\left\{\int^{z}_{z_{0}}\left[ -f'_{0}(\tilde{z})e^{-u(\tilde{z})} \right]d\tilde{z}\right\}e^{u(z)}.
\end{equation}
\begin{figure}[b]
\includegraphics[height=5cm,width=0.46\textwidth,angle=0]{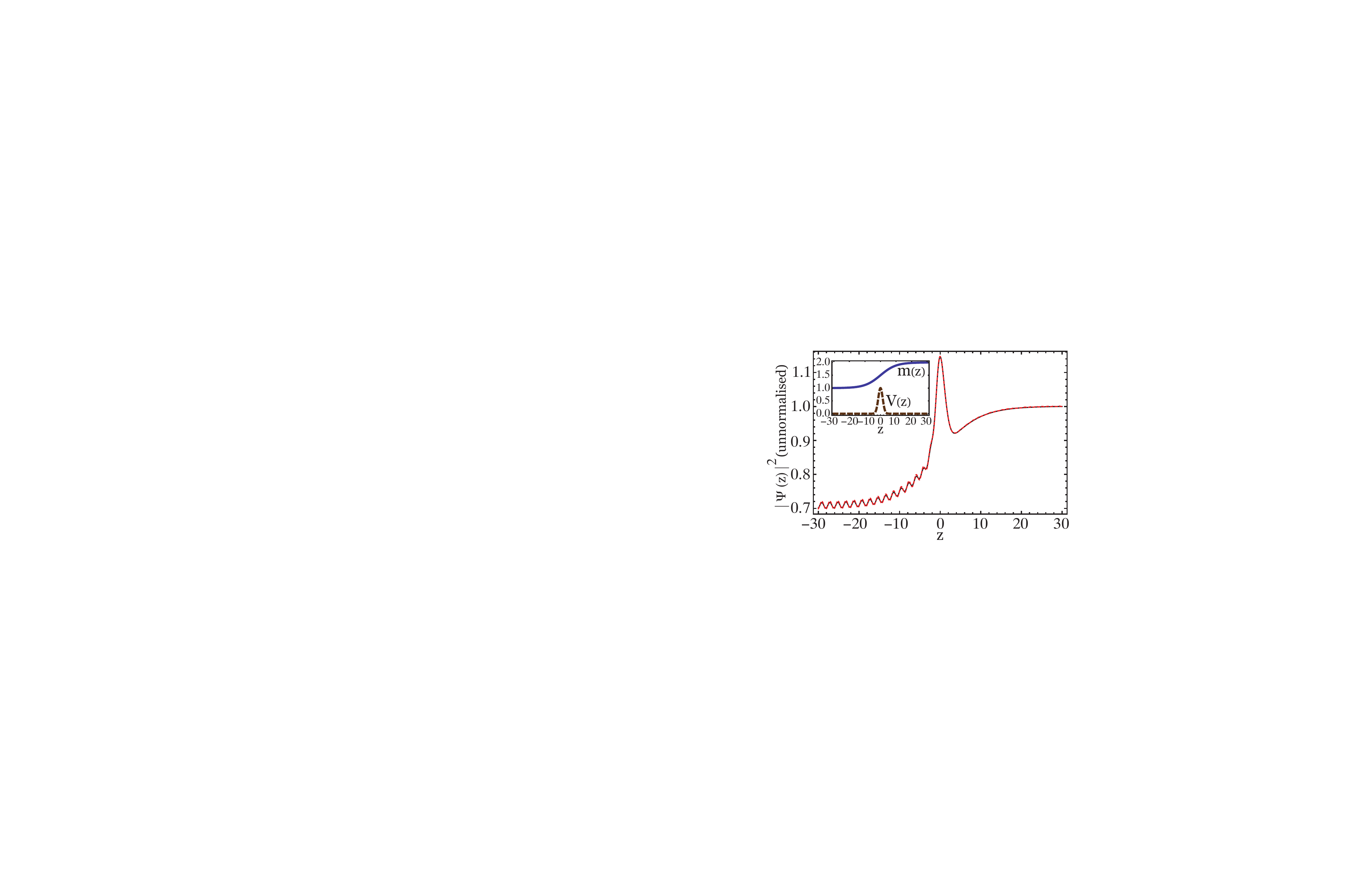}
\caption{(Color Online) Gaussian potential $V(z)=\text{exp}\left[-\beta^{2}z^{2}\right]$ and the hyperbolic mass variation $m(z)=(m_{1}+m_{2})/2+\left[(m_{1}-m_{2})/2\right]\text{tanh}(\alpha z)$ [see inset]. Plot of $|\Psi(z)|^2$ versus $z$ for the energy E=2.5. Dashed line represents the exact solution and solid line is the solution from
Eq.(\ref{518}). For numerical simulation $m_2=2$, $m_1=1$, $\alpha=0.1$ and $\gamma=0.5$} \label{fig1}
\end{figure}
The first order general solution of the Schr$\ddot{o}$dinger equation with position dependent mass Eq.(\ref{55}) is
\begin{eqnarray} \label{518}
    \Psi(z)&=A_{1}\text{exp}\left[\varphi(z)+i\theta(z)-\phi_{+}(z)\right]\nonumber \\
    &+A_{2}\text{exp}\left[\varphi(z)-i\theta(z)-\phi_{-}(z)\right],\nonumber \\
  \end{eqnarray}
where  $f_{\pm}(z), \varphi(z)$, $ \theta(z)$ is given by Eq.(\ref{510}), Eq.(\ref{512}a) and Eq.(\ref{512}b) respectively and $\phi_{\pm}(z)$ is defined as
\begin{equation}
\phi_{\pm}(z)=\int^{z}_{z_{0}}\left\{\int^{\tilde{z}}_{z_{0}}\left[ f'_{\pm}(\check{z})e^{\pm2i[\theta(\tilde{z})-\theta(\check{z})]}d\check{z}\right]\right\}d\tilde{z}.
\end{equation}
Using an iterative procedure our approach can be easily extended to include the next order corrections. Indeed, assuming $f_{2}(z)=f_{0}(z)+\epsilon_{1}(z)+\epsilon_{2}(z)$ we can easily obtain the equation for $\epsilon_{2}(z)$ as
\begin{equation}
\epsilon'_{2}(z)+2\left[f_{0}(z)+\epsilon_{1}(z)-\zeta(z)\right]\epsilon_{2}(z)+\epsilon^{2}_{1}(z)=0.
\end{equation}
In general, for $f_{n}(z)=f_{0}(z)+\sum^{n}_{j=1}\epsilon_{j}(z)$, we can obtain
\begin{equation}
\epsilon'_{n}(z)+2\left[f_{0}(z)+\sum^{n-1}_{j=1}\epsilon_{j}-\zeta(z)\right]\epsilon_{n}(z)+\epsilon^{2}_{n-1}(z)=0.
\end{equation}
All these equations can be solved exactly, thus we have obtained an analytical approximate solution for PDMSE.
\subsection{Application to scattering problem}
\noindent Let us now compare our approximate analytical solution with the exact solution for some physical problem. As an example, we study the scattering of a particle by a 1-D potential. A particle is propagating from left to right, during the propagation the effective mass of the particle varied due to its interaction with an ensemble of particles. For numerical simulations, we have used the scaling parameters given by Eq.(\ref{53}) and the corresponding PDMSE Eq.(\ref{57}). From elementary quantum mechanics we know that  due to the interaction of the particle with the potential, for $z\rightarrow +\infty$ the wave function for the particle is a plane wave while for $z\rightarrow -\infty$ it is the sum of the incident and the reflected plane waves. Thus we can write
\begin{figure}[t]
\includegraphics[height=5cm,width=0.46\textwidth,angle=0]{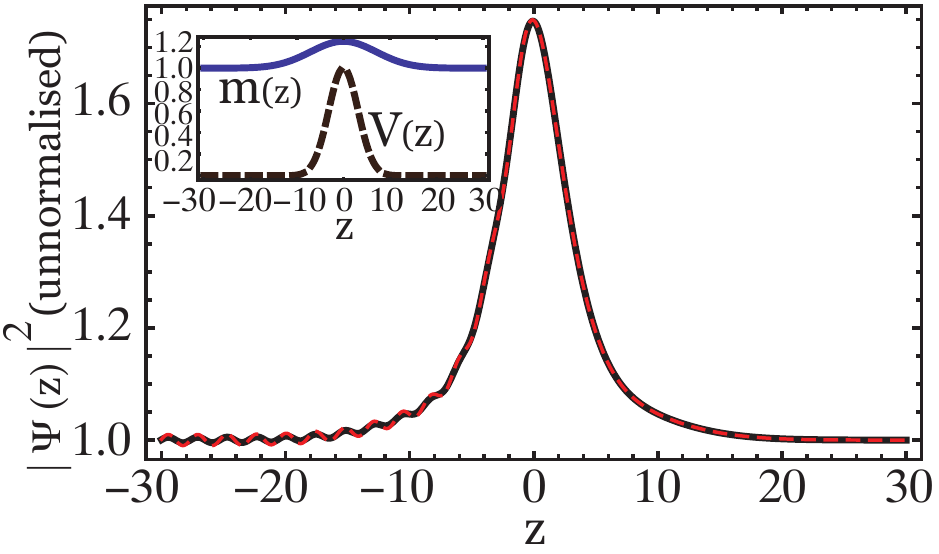}
\caption{(Color Online) Gaussian potential $V(z)=\exp\left[-\beta^{2}z^{2}\right]$ and the Gaussian-mass variation $m(z)=1+\text{exp}\left[-\delta^{2} z^{2}\right]$ [see inset].
Plot shows $|\Psi(z)|^2$ versus $z$ for the energy E=1.75. Dashed line represents the exact solution and solid line is the solution from
Eq.(\ref{518}). For numerical simulation $\beta=1/(2\sqrt{5}),\delta=0.1$ }
\end{figure}
\begin{equation}
\Psi(z) \rightarrow A \Psi_{i}(z)+B\Psi_{r}(z) \Big |_{z\rightarrow -\infty}
\end{equation}
Here the incident and the reflected plane wave have the form
\begin{subequations}
\begin{align}
\Psi_{i}(z)&\backsim \text{exp}\left[i\sqrt{E} z\right], \\
\Psi_{r}(z)&\backsim \text{exp}\left[-i\sqrt{E} z\right].
\end{align}
\end{subequations}
As a first example we will consider a Gaussian potential and the hyperbolic mass function of the for 
\begin{subequations}
\begin{align}
V(z)&=\text{exp}\left[-\beta^{2}z^{2}\right],\\
m(z)=\left[\frac{m_{1}+m_{2}}{2}\right]&+\left[\frac{m_{1}-m_{2}}{2}\right]\text{tanh}(\alpha z).
\end{align}
\end{subequations}
The mass $m(z)$ of the particle changes from $m_{1} \rightarrow m_{2}$. Fig. (1) shows the plot of the the probability density $|\Psi(z)|^{2}$ against $z$. The dashed line is the exact solution of the PDMSE Eq.(\ref{57}) while solid line is the analytical solution Eq.(\ref{518}).

Keeping the same form for the potential let us takes the mass function as $m(z)=1+\text{exp}\left[-\delta^{2} z^{2}\right]$. Here Fig. (2) shows the comparison plot for the numerical simulation and our analytical result. The smooth variation of the mass function from $m_{1}$ to $m_{2}$ [see Figs. (1,3) inset] can be controlled by the parameter $\alpha$ or $\delta$ as appropriate. Last example which we considered for the potential function is $V(z)=-\text{sech}(\gamma z)$ and the plot of the probability density (both numerical and analytical) is shown in Fig 3. Numerical simulation and analytical solution shown in Figs. (1,2,3) are nearly identical which suggests  that our method allows us to accurately obtain the wave function of a position dependent mass $m(x)$ particle scattered by a potential $\mathcal{V}(x)$. Our approach can be easily extended to find bound states also.
\subsection{Position Dependent Mass Schr$\ddot{\text{o}}$dinger Equation and Two-Level System }
 \noindent Let us consider a TLA with $|a\rangle$ and $|b\rangle$ representing the upper and the lower level states [see Fig. 4(a)].
 \begin{figure}[t]
\includegraphics[height=5cm,width=0.46\textwidth,angle=0]{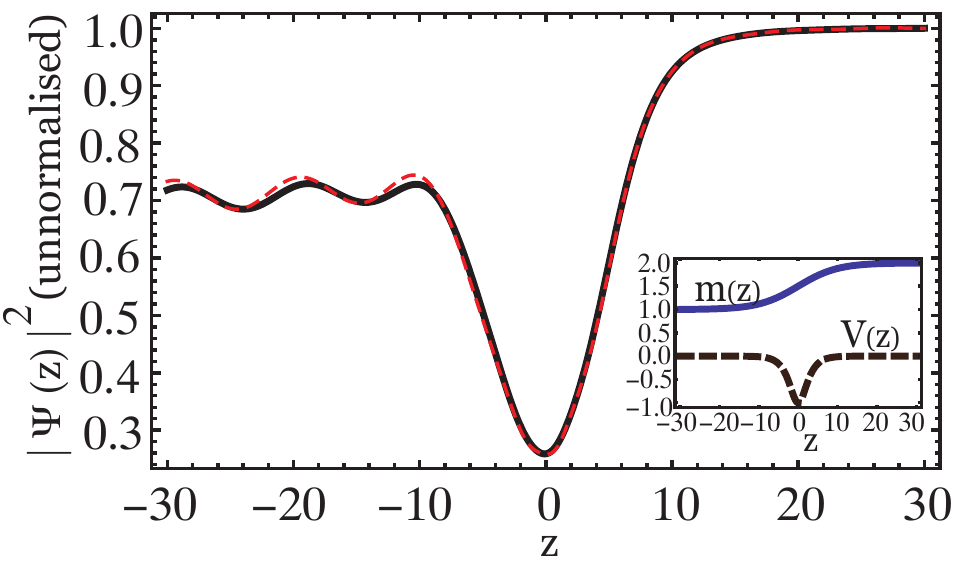}
\caption{(Color Online). Gaussian potential $V(z)=-\text{sech}(\gamma z)$ and the mass variation $m(z)=(m_{1}+m_{2})/2+\left[(m_{1}-m_{2})/2\right]\text{tanh}(\alpha z)$ [see inset].
$|\Psi(z)|^2$ versus $z$ for the energy E=0.1. Dashed line represents the exact solution and solid line is the solution from
Eq.(\ref{518}). For numerical simulation $m_2=2$, $m_1=1$, $\alpha=0.1$ and $\gamma=0.5$.}
\end{figure}
The equation of motion for the probability amplitudes $C_{a}$ and $C_{b}$ can be written as~\cite{I16}
\begin{subequations}\label{61}
\begin{align}
\dot{C}_{a}(t)&= i\Omega(t)\mbox{cos}(\nu t) e^{i\omega t}C_{b}(t),\label{second}\\
\dot{C}_{b}(t)&= i\Omega(t)\mbox{cos}(\nu t)e^{-i\omega t}C_{a}(t).
\end{align}
\end{subequations}
where $\hbar\omega$ is the energy difference between two levels, $\Omega(t)$ is the Rabi Frequency and considered real here. In invoking rotating-wave approximation we let $\mbox{cos}(\nu t)e^{\pm i\omega t}\rightarrow e^{\pm i\Delta t}/2$ where $\Delta=\omega-\nu$, is detuning from resonance.
\begin{subequations}\label{62}
\begin{align}
\dot{C}_{a}&= i\left[\Omega(t)/2\right] e^{i\Delta t}C_{b},\label{second}\\
\dot{C}_{b}&= i\left[\Omega(t)/2\right]e^{-i\Delta t}C_{a},
\end{align}
\end{subequations}
Coupled first order equations Eq.(\ref{62}) can be transformed to Riccati equation for $f=C_{a}(t)/C_{b}(t)$~\cite{I17,I171} or to the second order differential equation \cite{I18,I19}
\begin{equation}\label{63}
\ddot{C}_{a}(t)-\left[i\Delta + \frac{\dot{\Omega}(t)}{\Omega(t)}\right]\dot{C}_{a}(t)+\left[\frac{\Omega^{2}(t)}{4}\right]C_{a}(t)=0.
\end{equation}
In the case of resonance $\omega=\nu$ these equations Eq.(\ref{63}) transforms to
\begin{equation}\label{617}
\begin{split}
\ddot{C}_{a}(t)-\left[ \frac{\dot{\Omega}(t)}{\Omega(t)}\right]\dot{C}_{a}(t)+\left[\frac{\Omega^{2}(t)}{4}\right]C_{a}(t)=0.
\end{split}
\end{equation}
The exact  solution for the transformed equation is given as
\begin{equation}\label{618}
\begin{split}
C_{a}(t)=&A_{1}\text{exp}\left[i\int_{-\infty}^{t}  dt^{\prime }\Omega(t^{\prime })/2\right]+\\
&A_{2}\text{exp}\left[-i\int_{-\infty}^{t}  dt^{\prime }\Omega(t^{\prime })/2\right].
\end{split}
\end{equation}
\begin{figure}[t]
  \includegraphics[height=3.2cm,width=7.2cm,angle=0]{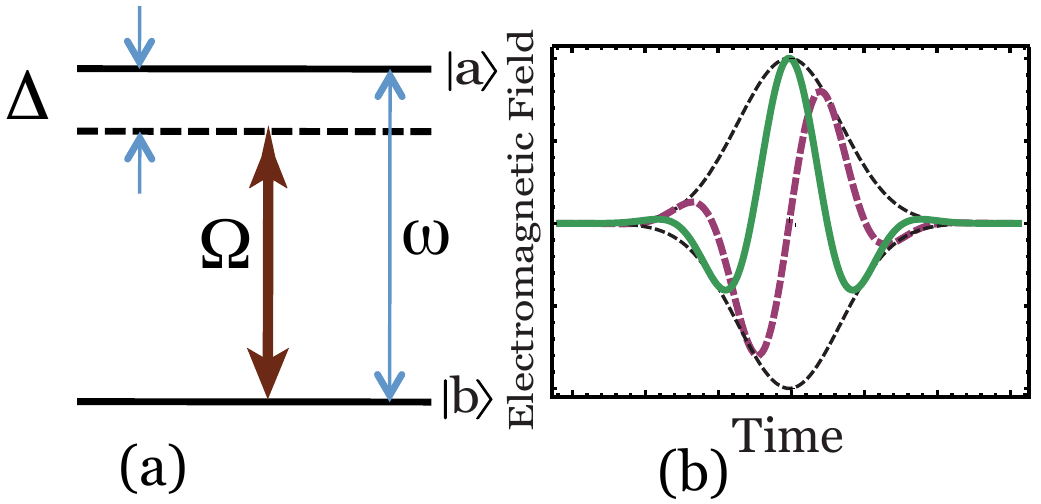}
  \caption{(Color Online) (a) Two-level atomic system, atomic transition frequency $\omega=\omega_{a}-\omega_{b}$, detuning $ \Delta = \omega -\nu$ and Rabi frequency $\Omega(t)$. (b) Few cycle sine (dashed line) and cosine (solid line) pulse with gaussian envelope.}
\end{figure}
\noindent Here $A_{1}$ and $A_{2}$ are constant which can be obtained using the initial condition of the problem. To see the connection between PDMSE and two-level system (TLS) let us consider that  the mass and potential function is related as
\begin{equation}\label{66}
\varrho(z)+V(z)=E.
\end{equation}
Subsequently from Eq.(\ref{55}) we get
\begin{equation}\label{67}
\left\{\frac{d^{2}}{dz^{2}}-\frac{\varrho'(z)}{\varrho(z)}\frac{d}{dz}+\rho(z)^{2}\right\}\Psi(z)=0.
\end{equation}
Solution to Eq.(\ref{67}) is given as
\begin{subequations}\label{68}
\begin{align}
\Psi(z) &=B_{1}e^{i\varphi(z)}+B_{2}e^{-i\varphi(z)},\label{second}\\
\varphi(z) &=\int{\varrho(z)dz}.
\end{align}
\end{subequations}
If we look Eq.(\ref{67}) with Eq.(\ref{617}) we see that they share the same structure. In one case it is the Rabi frequency $\Omega(t)$ which drives the TLA, while the variable mass $\varrho(z)$ plays a similar role for PDMSE. Thus the equivalence can be summarized as
\begin{subequations}\label{619}
\begin{align}
z&\leftrightarrow t,\label{second}\\
\Psi(z)&\leftrightarrow C_{a}(t), \label{third}\\
\varrho(z)&\leftrightarrow \Omega(t)/2.
\end{align}
\end{subequations}
Here we have assume that $\Omega(t)$ remains positive for all $t$. This structural analogy between the TLS and the PDMSE motivates us to find similar connections between two quantum-mechanical systems in two different areas of Physics.

To conclude, in this article we have studied the dynamics of position-dependent mass Schr$\ddot{o}$dinger equation using a recently developed technique (for constant mass Schr$\ddot{o}$dinger equation) to find analytical solution.  The  approximate solution is in excellent agreement with the numerical simulations [see Figs. 1,2,3]. We also investigated the structural analogy between coherent excitation of a TLA with a classical field in RWA and PDMSE. 

We thank M. O. Scully, L.V. Keldysh, M. S. Zubairy, Chia-Ren Hu, S.A.Chin, G. Agnolet for useful discussions. We also gratefully acknowledge the support from NSF Grant No. EEC-0540832 (MIRTHE ERC), Office of Naval Research (Grant Nos. N00014-09-1-0888 and N0001408- 1-0948), and the Robert A. Welch Foundation (Award No. A-1261) and partial support from the CRDF. P.K.J. also acknowledges the Robert A. Welch Foundation and HEEP Foundation for financial support.

\end{document}